\documentclass[aps,prd,preprint,showkeys]{revtex4}
\usepackage{graphicx}
\usepackage{subfigure}
\usepackage{hyperref}
\begin{document}
\draft
\title{Analyzing the anomalous dipole moment type couplings of heavy
quarks with FCNC interactions at the CLIC}
\author{A.~T. Tasci}
\email{atasci@kastamonu.edu.tr}
 %\affiliation{Department of
%Physics, Kastamonu University, 37100, Kastamonu, Turkey}
\author{A. Senol}
\email{asenol@kastamonu.edu.tr}
%\affiliation{Department of
%Physics, Kastamonu University, 37100, Kastamonu, Turkey}
\author{C. Verep}
%\email{cigdem_verep@hotmail.com}
 \affiliation{Department of
Physics, Kastamonu University, 37100, Kastamonu, Turkey}
%\pacs{12.60.-i, 13.66.Hk}
\begin{abstract}
In this study we examine both anomalous magnetic and dipole moment
type couplings of a heavy quark via its single production with
subsequent dominant Standard Model decay modes at the Compact Linear
Collider (CLIC). The signal and background cross sections are
analyzed for heavy quark masses 600 and 700 GeV. We make the
analysis to delimitate these couplings as well as to find the
attainable integrated luminosities for 3$\sigma$ observation limit.
\end{abstract}
\maketitle
\section{introduction}
Discovery of new particles performs a crucial role for physics
beyond the standard model (SM) and may play a milestone role in the
discovery of some open questions like the electroweak symmetry
breaking \cite{Holdom:1986rn,
Hill:1990ge,Elliott:1992xg,Hung:2010xh}, fermion mass spectrum
hierarchies and mixing angle in quark/lepton sectors
\cite{Holdom:2006mr,Hung:2007ak,Hung:2009ia,Hung:2009hy,Cakir:2009xi},
CP violation and flavor structure of standard theory
\cite{Hou:2010,BarShalom:2009sk,Buras:2010pi,Soni:2008bc,Eberhardt:2010bm,Soni:2010xh,Alok:2010zj}.
The precise determination of heavy quark properties may present the
existence of new physics. A heavy down-type quark ($b'$) with mass
less than 645 GeV and a up-type quark ($t'$) with mass less than 585
GeV \cite{TheATLAScollaboration:2013oha} are excluded at 95 \%
confidence level from proton-proton collisions at $\sqrt s=8$ TeV
ATLAS detector at the CERN Large Hadron Collider.

Searching for new sources of CP violation beyond the SM is an
attractive subject in particle physics since, it explains the
asymmetry between matter and anti-matter. CP violating anomalous
flavor changing neutral current (FCNC) $tcZ/tc\gamma$ couplings have
been considered in the literature before at hadron \cite{Han:1996ep}
and $e^-e^+$ \cite{Han:1998yr, Alan:2002fj} colliders. This type of
FCNC interactions offer an ideal place to search for new physics.
Due to the large mass values, heavy quarks have crucial advantage to
new interactions originating at a higher scale as in top quark
physics. Recently, anomalous FCNC $t$ quark couplings, such as $tqV$
($q=uc$, $V=\gamma,Z,g$), were experimentally restricted by some
collaborations. For instance, the upper limits observed from $tqg$
vertices by ATLAS Collaboration are $\kappa_{ugt}/\Lambda <6.9\times
10^{-3}$ TeV$^{-1}$ and $\kappa_{cgt}/\Lambda <1.6\times 10^{-2}$
TeV$^{-1}$ assuming only one coupling is kept nonzero
\cite{Aad:2012gd}, while D0 set limits as $\kappa_{tgu}/\Lambda
<0.013$ TeV$^{-1}$, $\kappa_{tgc}/\Lambda <0.057$ TeV$^{-1}$
\cite{Abazov:2010qk} and CDF set limits as $\kappa_{tug}/\Lambda
<0.018$ TeV$^{-1}$, $\kappa_{tcg}/\Lambda <0.069$ TeV$^{-1}$
\cite{Aaltonen:2008qr}. Recent observed upper limits on the coupling
strengths from CMS collaboration, which analysis both $gqt$ and
$Zqt$ vertices probed simultaneously, are $\kappa_{gut}/\Lambda
<0.10$ TeV$^{-1}$, $\kappa_{gct}/\Lambda <0.35$ TeV$^{-1}$,
$\kappa_{Zut}/\Lambda <0.45$ TeV$^{-1}$ and $\kappa_{Zct}/\Lambda
<2.27$ TeV$^{-1}$ \cite{Yazgan:2013pxa}.

Serious contributions can be expected for the production of the
heavy fermions, due to the anomalous magnetic moment type
interactions. Phenomenological studies with these anomalous effects
of these quarks have been performed at hadron colliders
\cite{Arik:2003vn,Arik:2002sg,Cakir:2009ib,Ciftci:2008tc,Sahin:2010wg,
Cakir:2012zz}, at ep colliders \cite{Alan:2003za,Ciftci:2009it} and
at linear colliders \cite{Senol:2011nm}. In this work, we study the
production of single heavy $t'$ quark at Compact Linear Collider
(CLIC) \cite{Linssen:2012hp} via both anomalous magnetic and dipole
moment type interactions. CLIC, a most popular proposed linear
collider on TeV scale, would complete the LHC results by performing
precision measurements to provide necessary information about some
parameters of heavy quarks. The aim of this study is to delimitate
the anomalous magnetic and dipole moment type couplings of $t'$
quark from a detailed signal and background analysis including Monte
Carlo simulation with the effects of initial state radiation (ISR)
and beamstrahlung (BS) in the $e^+e^-$ collisions.

\section{Single production and decay of $t'$ quark}

The interaction Lagrangian for $t'$ quark within the SM is given by
\begin{eqnarray}\label{L1}
L_s & = & -g_{e}Q_{t'}\overline{t}'\gamma^{\mu}t'A_{\mu}\nonumber \\
 &  & -g_{s}\overline{t}'T^{a}\gamma^{\mu}t'G_{\mu}^{a}\nonumber \\
 &  & -\frac{g_e}{2s_{W}c_{W}}\overline{t}'\gamma^{\mu}(g_{V}-g_{A}\gamma^{5})t'Z_{\mu}^{0}\nonumber \\
 &  & -\frac{g_e}{2\sqrt{2}s_{W}}V_{t'q_{i}}\overline{t}'\gamma^{\mu}(1-\gamma^{5})q_{i}W_{\mu}^{+}+h.c.\label{eq:1}
 \end{eqnarray}
where $A_{\mu}$, $G_{\mu}$, $Z_{\mu}^{0}$ and $W_{\mu}^+$ are the
vector fields for photon, gluon, $Z$ boson and $W$ boson,
respectively. $g_{e}$ is the electro-weak coupling constant and
$g_{s}$ is the strong coupling constant. $T^{a}$ are the Gell-Mann
matrices; $Q_{t'}$ is the electric charge of heavy quark $t'$.
$g_{V}$ and $g_{A}$ are the vector and axial-vector type couplings
of the neutral weak current with $t'$ quark, $\theta_W$ is the weak
mixing angle, $s_W=\sin\theta_W$ and $c_W=\cos\theta_W$. $V_{t'q}$
denotes the elements of extended 4$\times$4 CKM mixing matrix which
are constrained by flavor physics.

The anomalous magnetic and dipole moment type interactions among
heavy quark $t'$, ordinary quarks $q$, and the neutral gauge bosons
$V=\gamma,Z,g$ can be described by an effective Lagrangian which
contains the anomalous magnetic and dipole moment type couplings are
given by
\begin{eqnarray}\label{L2}
L'_{a} & = &
\sum_{q_{i}=u,c,t}Q_{q_{i}}\frac{g_{e}}{\Lambda}\overline{t}'\sigma_{\mu\nu}
(\kappa_{\gamma}^{q_{i}}-i\tilde\kappa_{\gamma}^{q_{i}}\gamma_5)q_{i}F^{\mu\nu}
\nonumber \\
 &  & +\sum_{q_{i}=u,c,t}\frac{g_{e}}{2\Lambda{s_W}{c_W}}\overline{t}'\sigma_{\mu\nu}
(\kappa_{Z}^{q_{i}}-i\tilde\kappa_{Z}^{q_{i}}\gamma_5)q_{i}Z^{\mu\nu}
\nonumber \\
 &  & +\sum_{q_{i}=u,c,t}\frac{g_{s}}{2\Lambda}\overline{t}'\sigma_{\mu\nu}
(\kappa_{g}^{q_{i}}-i\tilde\kappa_{g}^{q_{i}}\gamma_5)T^aq_{i}G_{a}^{\mu\nu}+h.c.
 \end{eqnarray}
where $F^{\mu\nu}$, $Z^{\mu\nu}$ and $G^{\mu\nu}$ are the field
strength tensors of the gauge bosons;
$\sigma_{\mu\nu}=i(\gamma_{\mu}\gamma_{\nu}-\gamma_{\nu}\gamma_{\mu})/2$;
$Q_{q_i}$ is the electric charge of the quark ($q$).
$\kappa_{\gamma}$($\tilde\kappa_{\gamma}$),
$\kappa_{Z}$($\tilde\kappa_{Z}$) and
$\kappa_{g}$($\tilde\kappa_{g}$) are the anomalous magnetic (dipole)
moment type couplings with photon, $Z$ boson and gluon,
respectively. Note that $\tilde\kappa$' s are CP violating,
$\Lambda$ is the cut off scale of new interactions and we assume
$\kappa_{\gamma}$=$\kappa_{Z}$=$\kappa_{g}$=$\kappa$ and
$\tilde\kappa_{\gamma}$=$\tilde\kappa_{Z}$=$\tilde\kappa_{g}$=$\tilde\kappa$.

\begin{figure}[hptb!]
\includegraphics[width=10cm]{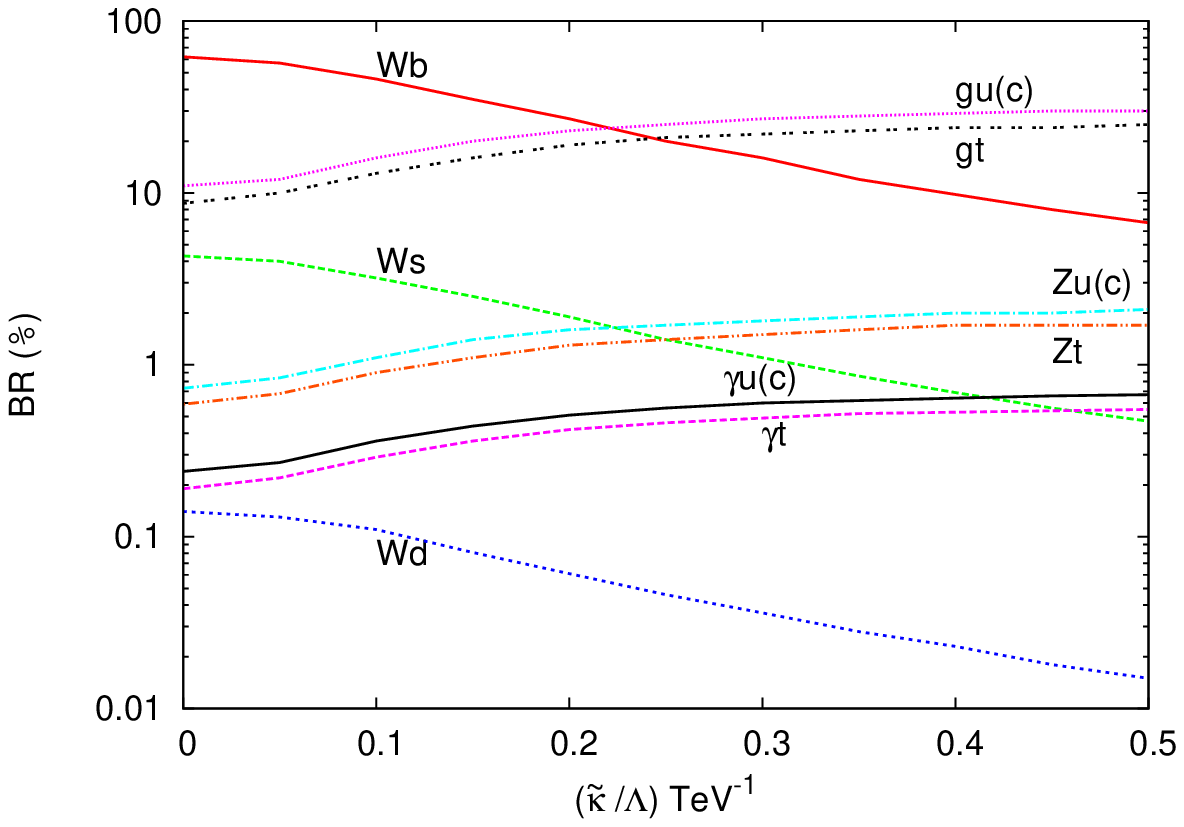} \caption{Branching ratios ($\%$) of all $t'$ decay channels depending on $\tilde{\kappa}/\Lambda$ for $m_{t'}=$700 GeV.}\label{br}
\end{figure}

\begin{figure}[hptb!]
\includegraphics[width=10cm]{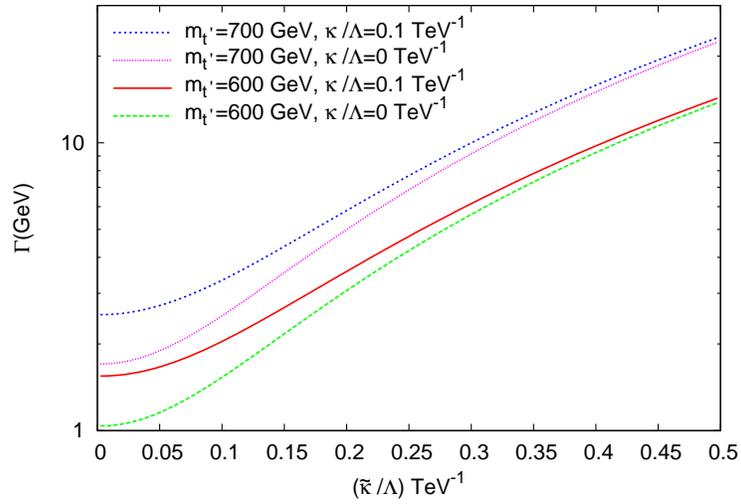} \caption{The total decay width of the $t'$ quark as function of $\tilde{\kappa}/\Lambda$.}\label{decay}
\end{figure}

CP-violating flavor changing neutral current processes within the SM
with the $b'$ and $t'$ quarks are analyzed by constructing and
employing global unique fit of the unitary $4\times 4$ CKM mass
mixing matrix at $m_{t'}$=600 and 700 GeV separately, in
Ref.\cite{Eilam:2009hz}. In our calculations we use this
parametrization for values of the $4\times 4$ CKM matrix elements
and we assume $m_{t'}>m_{b'}$ with a mass splitting of
$m_{t'}-m_{b'}\approx 50$ GeV. We implement the related interaction
vertices, given in the effective Lagrangian, into the tree level
event generator CompHEP package \cite{Pukhov:1999gg} for numerical
calculations. In Fig.~\ref{br}, branching ratios (BR) dependence on
$\tilde{\kappa}/\Lambda$ for SM decay channels ($Wd(s,b)$) and
anomalous decay channels ($Vu(c,t)$) of $t'$ quark which are
calculated by using Lagrangians (\ref{L1}) and (\ref{L2}) are given
for $m_{t'}=700$ GeV and $\kappa/\Lambda=0.1$ TeV$^{-1}$. As seen
from this figure, SM $t'$ decay channel, $t'\to W b$ is dominant for
$\tilde{\kappa}/\Lambda$ less than 0.2 changing around 27$\%$-63$\%$
BR. Total decay widths of $t'$ quark dependence on
$\tilde{\kappa}/\Lambda$ are given in Fig.~\ref{decay} for
$m_{t'}=$600 and 700 GeV with $\kappa/\Lambda=$0 and 0.1 TeV$^{-1}$.

\begin{figure}[hptb!]
\includegraphics[width=8cm]{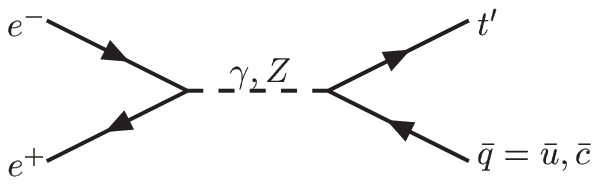}
\caption{Feynman diagram for anomalous single production of $t'$
quark in $e^+e^-$ collision.}\label{fig1}
\end{figure}

\begin{figure}[hptb!]
\includegraphics[width=10cm]{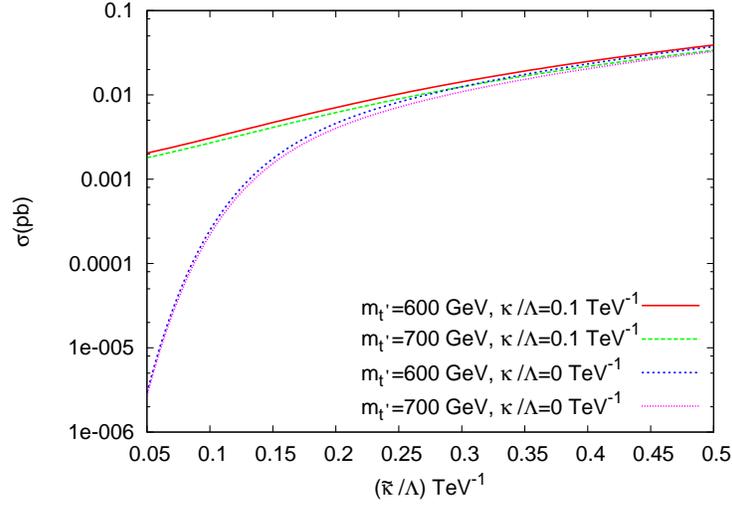} \caption{The
total cross sections at $\sqrt{s}=$3 TeV for the process $e^+e^-\to
t'\bar{q}~ (\bar{q}=\bar{u},\bar{c})$, as function of
$\tilde{\kappa}/\Lambda$.}\label{fig:cs1}
\end{figure}

The contributing tree level Feynman Diagram for the anomalous single
production of $t'$ quark in $e^+e^-$ collision is shown in
Fig.~\ref{fig1}. In Fig.~\ref{fig:cs1}, the total cross sections for
single production of $t'$ quark are plotted at collision center of
mass energy of 3 TeV with respect to $\tilde{\kappa}/\Lambda$ for
$m_{t'}=$600 and 700 GeV with $\kappa/\Lambda=$0 and 0.1 TeV$^{-1}$.
Initial state radiation (ISR) and beamstrahlung (BS) is a specific
feature of the linear colliders. We take the beam parameters for the
CLIC given in Table~\ref{tab2}, when calculating the ISR and BS
effects. Hereafter, in all our numerical calculations we take into
account ISR+BS effects.

\begin{table}
\caption{Main parameters of the CLIC. Here, N is the number of
particles in bunch. $\sigma_{x}$ and $\sigma_{y}$ are beam sizes,
$\sigma_{z}$ is the bunch length.}\label{tab2}
\begin{tabular}{lcc}
\hline Parameters     & CLIC\tabularnewline \hline
$E_{cm}(\sqrt{s})$ TeV    & $3$\tabularnewline
$L(10^{34}cm^{-2}s^{-1})$    & $5.9$\tabularnewline $N$$(10^{10})$ &
$0.372$\tabularnewline  $\sigma_{x}$ (nm)    & $45$\tabularnewline
$\sigma_{y}$ (nm)  & $1$\tabularnewline $\sigma_{z}$ ($\mu$m)  &
$44$\tabularnewline \hline
\end{tabular}
\end{table}
\section{Signal and Background Analysis}

The signal process of single production of $t'$ quark including the
dominant SM decay mode over anomalous decay is $e^+e^-\rightarrow t'
\bar{q_i}\rightarrow W^+ b~\bar{q_i}$ where,
$\bar{q_i}=\bar{u},\bar{c}$. The dominant source of SM background
process is $e^+e^-\rightarrow W^+ b~\bar{q_i}$ for the corresponding
signal processes.

\begin{figure}[hptb!]
\includegraphics[width=10cm]{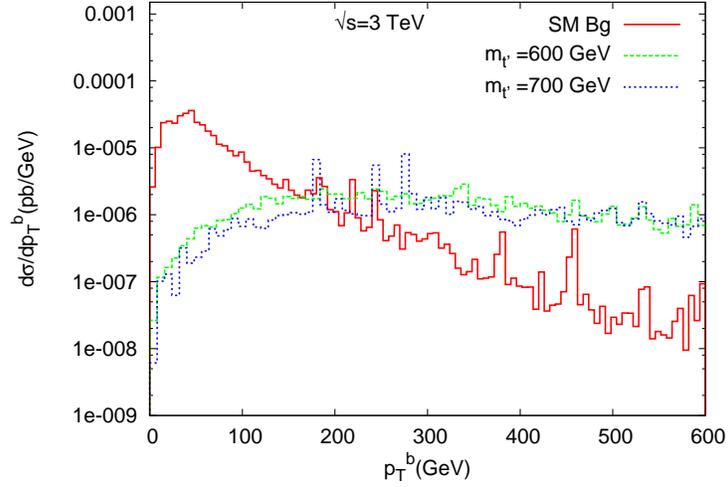}\caption{The
differential cross section depending on the transverse momentum of
the final state $b$ quark of process $e^+e^-\to W^{+}b~\bar{q}
(\bar{q}=\bar{u},\bar{c})$ for SM background (solid line) and signal
with different mass values of $t'$ quarks at $\sqrt{s}=$3
TeV.}\label{fig2}
\end{figure}

In the transverse momentum, rapidity and invariant mass
distributions analysis we assume
$\tilde{\kappa}/\Lambda=\kappa/\Lambda=$0.1 TeV$^{-1}$. In
Fig.~\ref{fig2}, the transverse momentum ($p_T$) distributions of
the final state $b$ quark for signal and background are shown for
CLIC energy. We applied a $p_T$ cut of $p_T> 50$ GeV to reduce the
background, comparing the signal $p_T$ distribution of $b$ quark
with that of the corresponding background.

\begin{figure}[hptb!]
\includegraphics[width=10cm]{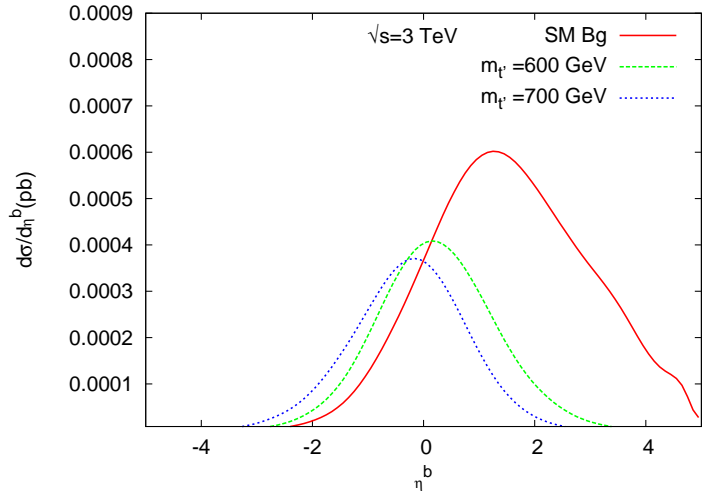} \caption{The
rapidity distribution of the final state $b$ quark at $\sqrt{s}=$3
TeV for the process $e^+e^-\to W^+b\bar{q}
~(\bar{q}=\bar{u},\bar{c})$.}\label{fig3}
\end{figure}

In Fig.~\ref{fig3}, we plot the rapidity distributions of final
state $b$ quark in signal and background processes. According to
these figures, the cut $|\eta^b|<2.5$ can be applied to suppress the
background while the signal remains almost unchanged.

\begin{figure}[hptb!]
\includegraphics[width=10cm]{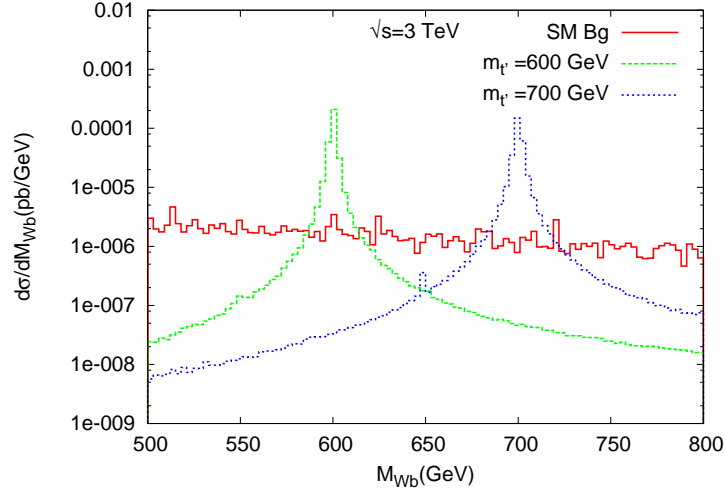} \caption{The invariant
mass distribution of the final state $Wb$ system for SM background
(solid line) and signal from $t'$ decay for  $m_{t'}=600$ GeV
(dashed line) and $m_{t'}=700$ GeV (dot-dashed line) at $\sqrt{s}=$3
TeV.}\label{fig4}
\end{figure}

In Fig.~\ref{fig4}, the invariant mass distributions for the $W^+b$
system in the final state are plotted. From these figures, we can
see that the signal has a peak around mass of $t'$ quark over the
background.

In Table \ref{ss3}, we calculate cross sections the signal and
background and the statistical significance ($SS$) to discuss the
observability of 600 and 700 GeV $t'$ quark for
$\tilde{\kappa}/\Lambda=$0.1 and 0.01 TeV$^{-1}$ by taking
$\kappa/\Lambda=$0 TeV$^{-1}$ at CLIC. The $SS$ of the signal are
obtained by using the formula,
\begin{eqnarray*}
SS&=&(\sigma_{S}/\sqrt{\sigma_{S}+\sigma_{B}})\sqrt{BR(W\to
l\nu_l)\cdot L_{int}}.
\end{eqnarray*}
where $\sigma_{S}$ is the the signal and $\sigma_{B}$ is the
background cross sections for the $e^+e^-\rightarrow t'
\bar{q_i}\rightarrow W^+ b~\bar{q_i}$ process, respectively and
$l=e,\mu$. We take into account finite energy resolution of the
detectors for realistic analysis. In our numerical calculations we
use the mass bin width $\Delta m=max(2\Gamma,\delta m)$ to count
signal and background events with the mass resolution $\delta m$.
The mentioned $p_T$ and $\eta$ cuts are applied assuming the
integrated luminosity given in Table~\ref{tab2}.
\begin{table}[hptb!]
\caption{The signal and background cross sections and signal
 Statistical Significance ($SS$) by taking $\kappa/\Lambda=0$ TeV$^{-1}$ for the CLIC at $\sqrt s=3$ TeV with integrated
 luminosity of 5.9$\times10^5$ pb$^{-1}$.}\label{ss3}
\begin{tabular}{|l|l|l|l|l|l|l|}

  \cline{2-7}
\multicolumn{1}{l|}{}&  \multicolumn{3}{ c| }{$\tilde{\kappa}/\Lambda=0.1$ TeV$^{-1}$}& \multicolumn{3}{ c| }{$\tilde{\kappa}/\Lambda=0.01$ TeV$^{-1}$} \\
 \hline
  $m_{t'}$(GeV) & $\sigma_S$(fb) & $\sigma_B$(fb) & $SS$ & $\sigma_S$(fb) & $\sigma_B$(fb) & $SS$\\\hline
  600 & $3.01$ & $8.92\times 10^{-3}$ & $19.72$ & $1.50\times 10^{-2}$ & $8.92\times 10^{-3}$ & $1.10$ \\
  700 & $2.63$ & $1.14\times 10^{-2}$ & $18.43$ & $1.31\times 10^{-2}$ & $1.14\times 10^{-2}$ & $0.95$ \\
  \hline
\end{tabular}
\end{table}

\begin{figure}[hptb!]
\includegraphics[width=10cm]{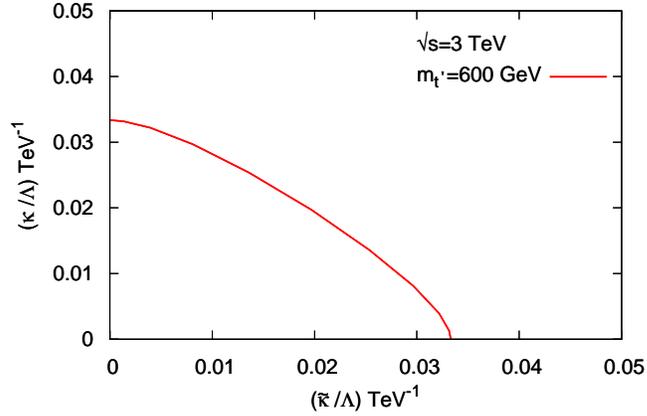}
\caption{The 3$\sigma$ contour plot for the anomalous couplings
reachable at $\sqrt{s}=$3 TeV with $L_{int}$=5.9$\times10^5$
pb$^{-1}$ for $m_{t'}$=600 GeV.}\label{cp1}
\end{figure}

After this point, we will focus on limiting the anomalous magnetic
and dipole moment type couplings. Firstly, in Fig.~\ref{cp1}, we
present the 3$\sigma$ contour plot for $\tilde\kappa/\Lambda$ -
$\kappa/\Lambda$ plane at $\sqrt s=3$ TeV with $m_{t'}$=600 GeV.
According to these figures, the lower limits of $\kappa/\Lambda$ and
$\tilde\kappa/\Lambda$ are about 0.033 TeV$^{-1}$ at the CLIC
energy.

\begin{figure}[hptb!]
 \subfigure[]{\includegraphics[width=8cm]{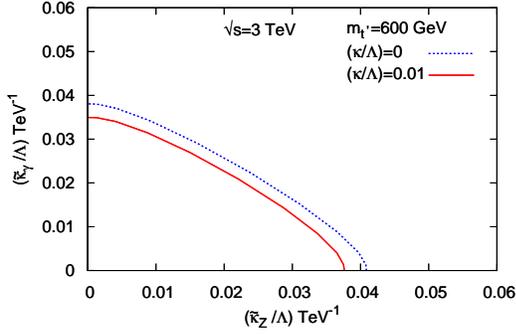}}
\hspace{0.1cm}
\subfigure[]{\includegraphics[width=8cm]{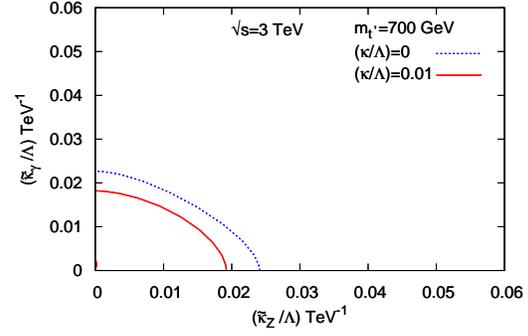}}
\caption{The 3$\sigma$ contour plot for the anomalous dipole moment
type couplings reachable at (a) $m_{t'}$=600 GeV and (b)
$m_{t'}$=700 GeV.}\label{cp2}
\end{figure}

To analyze the case of
$\tilde\kappa_{\gamma}/\Lambda\neq\tilde\kappa_Z/\Lambda$, the
3$\sigma$ contour plots for the anomalous couplings in the
$\tilde\kappa_Z/\Lambda-\tilde\kappa_{\gamma}/\Lambda$ plane are
presented in Fig.~\ref{cp2} at $\sqrt s=3$ TeV with a) $m_{t'}$=600
GeV and b) $m_{t'}$=700 GeV by taking into account different values
of $\kappa/\Lambda$. According to these figures the lower limits of
$\tilde\kappa_{\gamma}/\Lambda$ and $\tilde\kappa_Z/\Lambda$ are
about 0.038 TeV$^{-1}$ for $m_{t'}$=600 GeV and 0.019 TeV$^{-1}$ for
$m_{t'}$=700 GeV with $\kappa/\Lambda$=0.01 TeV$^{-1}$. In
Figs.~\ref{cp1} and \ref{cp2}, allowed parameter space area of $t'$
quark is above the lines.

\begin{figure}[hptb!]
 \subfigure[]{\includegraphics[width=8cm]{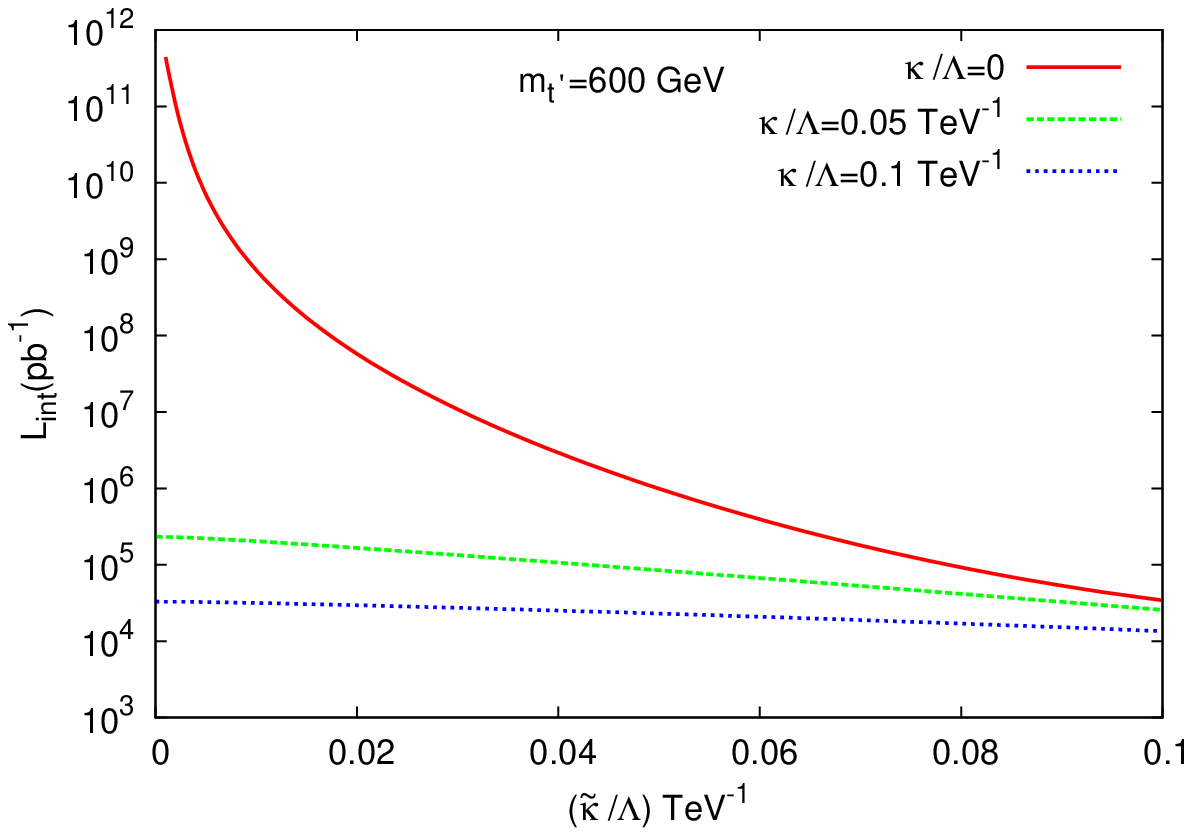}}
\hspace{0.1cm}
\subfigure[]{\includegraphics[width=8cm]{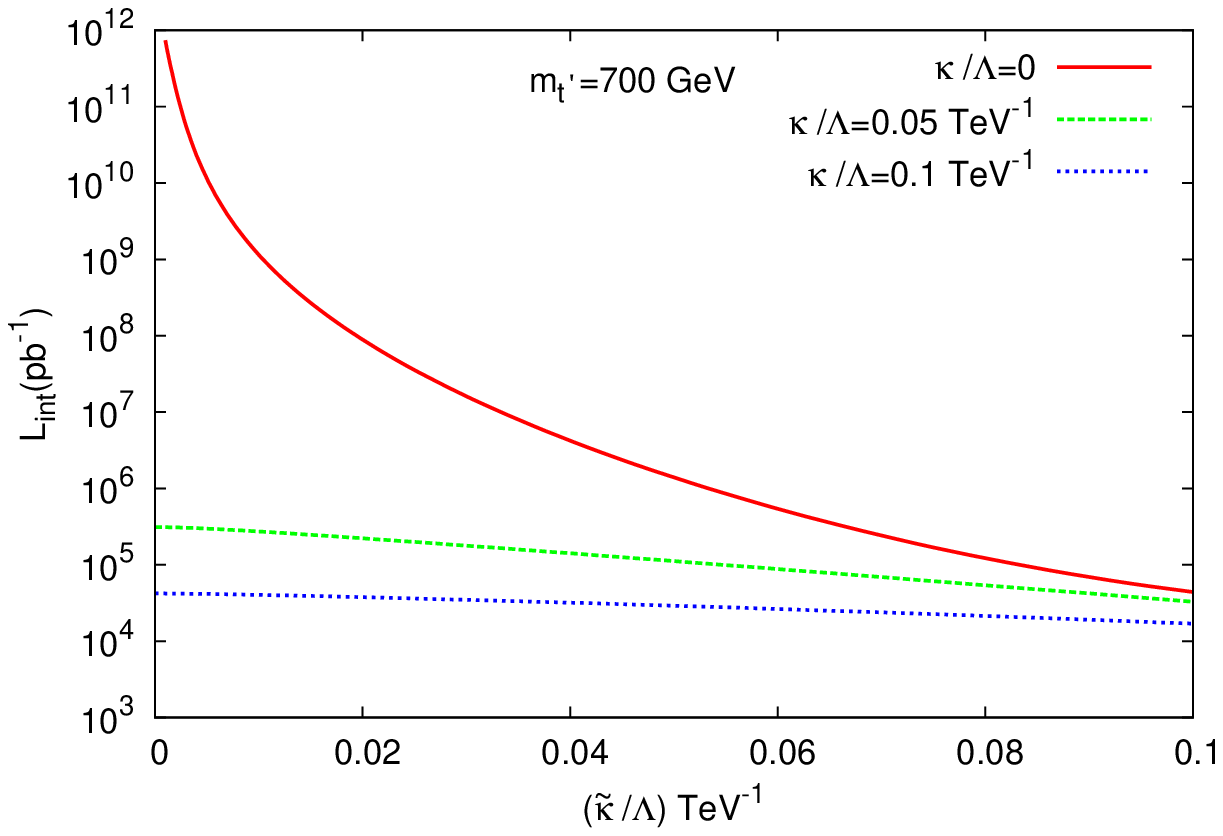}}
\caption{The attainable integrated luminosity for 3$\sigma$
observation limit for (a) $m_{t'}$=600 GeV (b) $m_{t'}$=700 GeV for
$\sqrt{s}=$3 TeV depending on anomalous dipole moment type
couplings.}\label{lp}
\end{figure}

We plot the lowest necessary luminosities with 3$\sigma$ observation
limits for (a) $m_{t'}$=600 GeV and (b) $m_{t'}$=700 GeV at
$\sqrt{s}=$3 TeV depending on anomalous couplings in Fig.~\ref{lp}.
In the case of $\kappa/\Lambda$=$\tilde\kappa/\Lambda$=0.1
TeV$^{-1}$, it is seen that from these figures, $t'$ quarks with
masses 600 and 700 GeV can be observed at 3$\sigma$ observation
limit with lowest integrated luminosity about at the order of $10^4$
pb$^{-1}$ at CLIC.

\section{Conclusion}
The anomalous FCNC interactions of heavy quarks could be important
for some parameter regions due to the expected large masses. The
sensitivity to the anomalous couplings ($\kappa$, $\tilde{\kappa}$)
and ($\tilde\kappa_{\gamma}$, $\tilde{\kappa_Z}$) can be obtained
for $m_{t'}$=600 GeV about (0.033, 0.033) and (0.035, 0.038) for
$\kappa$=0.01, for $m_{t'}$=700 GeV ($\tilde\kappa_{\gamma}$,
$\tilde{\kappa_Z}$) values can be obtained about (0.019, 0.0195) for
$\kappa$=0.01 with $\Lambda$= 1 TeV. We also find the lowest
necessary luminosity limit values at the order of $10^4$ pb$^{-1}$
for CLIC.

\begin{acknowledgments} This research has been partially supported by Kastamonu
University Scientific Research Projects Coordination Department
under the grant No. KUBAP-03/2012-01.
\end{acknowledgments}

\newpage

\begin{thebibliography}{99}
%\cite{}
 \bibitem{Holdom:1986rn}
  B.~Holdom,
  %``Heavy Quarks And Electroweak Symmetry Breaking,''
  Phys.\ Rev.\ Lett.\  {\bf 57}, 2496 (1986)
  [Erratum-ibid.\  {\bf 58}, 177 (1987)].
  %%CITATION = PRLTA,57,2496;%%
%\cite{}
\bibitem{Hill:1990ge}
  C.~T.~Hill, M.~A.~Luty and E.~A.~Paschos,
  %``Electroweak symmetry breaking by fourth generation condensates and the
  %neutrino spectrum,''
  Phys.\ Rev.\  D {\bf 43}, 3011 (1991).
  %%CITATION = PHRVA,D43,3011;%%
%\cite{}
\bibitem{Elliott:1992xg}
  T.~Elliott and S.~F.~King,
  %``Heavy quark condensates from dynamically broken flavor symmetry,''
  Phys.\ Lett.\  B {\bf 283}, 371 (1992).
  %%CITATION = PHLTA,B283,371;%%
%\cite{}
\bibitem{Hung:2010xh}
  P.~Q.~Hung and C.~Xiong,
  %``Dynamical Electroweak Symmetry Breaking with a Heavy Fourth Generation,''
  Nucl.\ Phys.\  B {\bf 848}, 288 (2011).
  %[arXiv:1012.4479 [hep-ph]].
  %%CITATION = NUPHA,B848,288;%%

\bibitem{Holdom:2006mr}
  B.~Holdom,
  %``The discovery of the fourth family at the LHC: What if?,''
  JHEP {\bf 0608}, 076 (2006).
  %[arXiv:hep-ph/0606146].
  %%CITATION = JHEPA,0608,076;%%
%\cite{Hung:2007ak}
\bibitem{Hung:2007ak}
  P.~Q.~Hung and M.~Sher,
  %``Experimental constraints on fourth generation quark masses,''
  Phys.\ Rev.\  D {\bf 77}, 037302 (2008).
  %[arXiv:0711.4353 [hep-ph]].
  %%CITATION = PHRVA,D77,037302;%%

%%\cite{}
\bibitem{Hung:2009ia}
  P.~Q.~Hung, C.~Xiong,
  %``Implication of a Quasi Fixed Point with a Heavy Fourth Generation: The emergence of a TeV-scale physical cutoff,''
  Phys.\ Lett.\  {\bf B694}, 430-434 (2011).
  %[arXiv:0911.3892 [hep-ph]].
%\cite{}
\bibitem{Hung:2009hy}
  P.~Q.~Hung and C.~Xiong,
  %``Renormalization Group Fixed Point with a Fourth Generation: Higgs-induced
  %Bound States and Condensates,''
  Nucl.\ Phys.\  B {\bf 847}, 160 (2011).
  %[arXiv:0911.3890 [hep-ph]].
  %%CITATION = NUPHA,B847,160;%%

  %\cite{}
\bibitem{Cakir:2009xi}
  O.~Cakir, A.~Senol and A.~T.~Tasci,
  %``Single Production of Fourth Family t-prime Quarks at LHeC,''
  Europhys.\ Lett.\  {\bf 88}, 11002 (2009).
  %[arXiv:0905.4347 [hep-ph]].
  %%CITATION = EULEE,88,11002;%%

\bibitem{Hou:2010}
  W.~S.~Hou and C.~Y.~Ma,
  %``Flavor and CP Violation with Fourth Generations Revisited,''
  Phys.\ Rev.\  D {\bf 82}, 036002 (2010).
  %[arXiv:1004.2186 [hep-ph]].
  %%CITATION = PHRVA,D82,036002;%%
  %\cite{}
  \bibitem{BarShalom:2009sk}
  S.~Bar-Shalom, D.~Oaknin and A.~Soni,
  %``Extended Friedberg Lee hidden symmetries, quark masses and CP-violation
  %with four generations,''
  Phys.\ Rev.\  D {\bf 80}, 015011 (2009).
  %[arXiv:0904.1341 [hep-ph]].
  %%CITATION = PHRVA,D80,015011;%%

\bibitem{Buras:2010pi}
  A.~J.~Buras, B.~Duling, T.~Feldmann, T.~Heidsieck, C.~Promberger and S.~Recksiegel,
  %``Patterns of Flavour Violation in the Presence of a Fourth Generation of
  %Quarks and Leptons,''
  JHEP {\bf 1009}, 106 (2010).
  %[arXiv:1002.2126 [hep-ph]].
  %%CITATION = JHEPA,1009,106;%%
%\cite{}
\bibitem{Soni:2008bc}
  A.~Soni, A.~K.~Alok, A.~Giri, R.~Mohanta and S.~Nandi,
  %``The Fourth family: A Natural explanation for the observed pattern of
  %anomalies in $B^-$ CP asymmetries,''
  Phys.\ Lett.\  B {\bf 683}, 302 (2010).
  %[arXiv:0807.1971 [hep-ph]].
  %%CITATION = PHLTA,B683,302;%%
  %\cite{}
\bibitem{Eberhardt:2010bm}
  O.~Eberhardt, A.~Lenz and J.~Rohrwild,
  %``Less space for a new family of fermions,''
  Phys.\ Rev.\  D {\bf 82}, 095006 (2010).
  %[arXiv:1005.3505 [hep-ph]].
  %%CITATION = PHRVA,D82,095006;%%

\bibitem{Soni:2010xh}
A.~Soni, A.~K.~Alok, A.~Giri, R.~Mohanta and S.~Nandi,
%``SM with four generations: Selected implications for rare B and K decays,''
Phys.\ Rev.\  D {\bf 82}, 033009 (2010).
%[arXiv:1002.0595 [hep-ph]].
%%CITATION = PHRVA,D82,033009;%%


\bibitem{Alok:2010zj}
 A.~K.~Alok, A.~Dighe and D.~London,
 %``Constraints on the Four-Generation Quark Mixing Matrix from a Fit to
 %Flavor-Physics Data,''
 Phys.\ Rev.\  D {\bf 83}, 073008 (2011).
 %[arXiv:1011.2634 [hep-ph]].
 %%CITATION = PHRVA,D83,073008;%%

%\cite{TheATLAScollaboration:2013oha}
\bibitem{TheATLAScollaboration:2013oha}
  The ATLAS collaboration,
  %``Search for pair production of new heavy quarks that decay to a $\mathbf{Z}$ boson and a third generation quark in $\mathbf{pp}$ collisions at $\mathbf{\sqrt{s}=8}$ TeV with the ATLAS detector,''
  ATLAS-CONF-2013-056.

%\cite{Han:1996ep}
\bibitem{Han:1996ep}
  T.~Han, K.~Whisnant, B.~L.~Young and X.~Zhang,
  %``Top quark decay via the anomalous coupling $\bar{t} c \gamma$ at hadron colliders,''
  Phys.\ Rev.\ D {\bf 55}, 7241 (1997).
  %  [hep-ph/9603247].  %%CITATION = HEP-PH/9603247;%%
%\cite{Han:1998yr}
\bibitem{Han:1998yr}
  T.~Han and J.~L.~Hewett,
  %``Top charm associated production in high-energy $e^{+} e^{-}$ collisions,''
  Phys.\ Rev.\ D {\bf 60}, 074015 (1999).
  %  [hep-ph/9811237].  %%CITATION = HEP-PH/9811237;%%
%\cite{Alan:2002fj}
\bibitem{Alan:2002fj}
  A.~T.~Alan, A.~Senol and A.~T.~Tasci,
  %``CP violating asymmetries in the flavor changing single top quark production,''
  J.\ Phys.\ G {\bf 29}, 279 (2003).
  % [hep-ph/0205195].  %%CITATION = HEP-PH/0205195;%%

%\cite{Aad:2012gd}
\bibitem{Aad:2012gd}
  G.~Aad {\it et al.}  [ATLAS Collaboration],
  %``Search for FCNC single top-quark production at $\sqrt{s}=7$ TeV with the ATLAS detector,''
  Phys.\ Lett.\ B {\bf 712}, 351 (2012).
%  [arXiv:1203.0529 [hep-ex]].
  %%CITATION = ARXIV:1203.0529;%%
  %44 citations counted in INSPIRE as of 31 Jan 2014
 %\cite{Abazov:2010qk}
\bibitem{Abazov:2010qk}
  V.~M.~Abazov {\it et al.}  [D0 Collaboration],
  %``Search for flavor changing neutral currents via quark-gluon couplings in single top quark production using 2.3 fb$^{-1}$ of $p\bar{p}$ collisions,''
  Phys.\ Lett.\ B {\bf 693}, 81 (2010).
 % [arXiv:1006.3575 [hep-ex]].
  %%CITATION = ARXIV:1006.3575;%%
%\cite{Aaltonen:2008qr}
\bibitem{Aaltonen:2008qr}
  T.~Aaltonen {\it et al.}  [CDF Collaboration],
  %``Search for top-quark production via flavor-changing neutral currents in W+1 jet events at CDF,''
  Phys.\ Rev.\ Lett.\  {\bf 102}, 151801 (2009).

%\cite{Yazgan:2013pxa}
\bibitem{Yazgan:2013pxa}
  E.~Yazgan [ for the ATLAS and CDF and CMS and D0 Collaborations],
  %``Flavor changing neutral currents in top quark production and decay,''
  arXiv:1312.5435 [hep-ex].
  %%CITATION = ARXIV:1312.5435;%%

\bibitem{Arik:2003vn}
  E.~Arik, O.~Cakir, S.~Sultansoy,
  %``Search for anomalous single production of the fourth SM family quark decaying into a light scalar,''
  Europhys.\ Lett.\  {\bf 62}, 332-335 (2003).
  %[hep-ph/0309041].
  %\cite{}

\bibitem{Arik:2002sg}
  E.~Arik, O.~Cakir, S.~Sultansoy,
  %``Anomalous single production of the fourth SM family quarks at Tevatron,''
  Phys.\ Rev.\  {\bf D67}, 035002 (2003).
  %[hep-ph/0208033].

  %\cite{}
\bibitem{Cakir:2009ib}
  I.~T.~Cakir, H.~Duran Yildiz, O.~Cakir {\it et al.},
  %``Anomalous resonant production of the fourth family up type quarks at the LHC,''
  Phys.\ Rev.\  {\bf D80}, 095009 (2009).
  %[arXiv:0908.0123 [hep-ph]].

  %\cite{}
\bibitem{Ciftci:2008tc}
  R.~Ciftci,
  %``Anomalous Single Production of the Fourth Generation Quarks at the LHC,''
  Phys.\ Rev.\  {\bf D78}, 075018 (2008).
  %[arXiv:0807.4291 [hep-ph]].

%\cite{}
\bibitem{Sahin:2010wg}
  M.~Sahin, S.~Sultansoy, S.~Turkoz,
  %``A Search for the Fourth SM Family Quarks through Anomalous Decays,''
  Phys.\ Rev.\  {\bf D82}, 051503 (2010).
  %[arXiv:1005.4538 [hep-ph]].
  %\cite{Cakir:2012zz}
\bibitem{Cakir:2012zz}
  O.~Cakir, I.~T.~Cakir, A.~Senol and A.~T.~Tasci,
  %``Anomalous single production of fourth family up-type quark associated with neutral gauge bosons at the LHC,''
  J.\ Phys.\ G {\bf 39}, 055005 (2012).
   %%CITATION = JPHGB,G39,055005;%%

\bibitem{Alan:2003za}
  A.~T.~Alan, A.~Senol, O.~Cakir,
  %``Anomalous production of fourth family up quarks at future lepton hadron colliders,''
  Europhys.\ Lett.\  {\bf 66}, 657-660 (2004).
  %[hep-ph/0312021].
\bibitem{Ciftci:2009it}
  R.~Ciftci and A.~K.~Ciftci,
  %``A Comperative Study of the Anomalous Single Production of the Fourth
  %Generation Quarks at ep and gamma-p Colliders,''
  arXiv:0904.4489 [hep-ph].
  %%CITATION = ARXIV:0904.4489;%%
%\cite{Senol:2011nm}
\bibitem{Senol:2011nm}
  A.~Senol, A.~T.~Tasci and F.~Ustabas,
  %``Anomalous single production of fourth generation $t'$ quarks at ILC and CLIC,''
  Nucl.\ Phys.\ B {\bf 851}, 289 (2011).
 % [arXiv:1104.5316 [hep-ph]].  %%CITATION = ARXIV:1104.5316;%%

 %\cite{Linssen:2012hp}
\bibitem{Linssen:2012hp}
  L.~Linssen, A.~Miyamoto, M.~Stanitzki and H.~Weerts,
  ``Physics and Detectors at CLIC: CLIC Conceptual Design Report,''
  arXiv:1202.5940 [physics.ins-det].  %%CITATION = ARXIV:1202.5940;%%


%\cite{Eilam:2009hz}
\bibitem{Eilam:2009hz}
  G.~Eilam, B.~Melic and J.~Trampetic,
  %``CP violation and the 4th generation,''
  Phys.\ Rev.\ D {\bf 80}, 116003 (2009).
  \bibitem{Pukhov:1999gg} A.~Pukhov \textit{et al.}, %``CompHEP: A package for evaluation of Feynman diagrams and integration  over
 %multi-particle phase space. User's manual for version 33,''
 arXiv:hep-ph/9908288.


  \end{thebibliography}
\end{document}